\documentclass[unnumsec,webpdf,contemporary,large]{oup-authoring-template}%

\graphicspath{{Fig/}}

\usepackage{amsmath,amssymb}
\usepackage{float}
\usepackage{algorithm}
\usepackage{algpseudocode}
\usepackage{booktabs}
\usepackage{xurl}
\usepackage{microtype}
\usepackage{caption}    

\usepackage{amsmath,amssymb}

\allowdisplaybreaks

\theoremstyle{thmstyleone}%
\theoremstyle{thmstyletwo}%
\theoremstyle{thmstylethree}%

\begin{document}

\journaltitle{Journal Title Here}
\DOI{DOI added during production}
\copyrightyear{YEAR}
\pubyear{YEAR}
\vol{XX}
\issue{x}
\access{Published: Date added during production}
\appnotes{Paper}

\firstpage{1}


\title[Short Article Title]{A Joint Survival Modeling and Therapy Knowledge Graph Framework to Characterize Opioid Use Disorder Trajectories}

\author[1,$\ast$]{Mengman Wei\ORCID{0009-0008-0458-7140}}
\author[2]{Qian Peng}

\address[1]{\orgdiv{Department of Neuroscience}, \orgname{The Scripps Research Institute}, \orgaddress{\street{10550 N Torrey Pines Rd, La Jolla}, \postcode{92037}, \state{CA}, \country{U.S.}}}

\corresp[$\ast$]{Mengman Wei. \href{email:email-id.com}{mwei@scripps.edu}}

\received{Date}{0}{Year}
\revised{Date}{0}{Year}
\accepted{Date}{0}{Year}



\abstract{%
Motivation: Opioid use disorder (OUD) often emerges following prescription opioid exposure and follows a dynamic clinical course characterized by onset, remission, and relapse. Large-scale cohorts linking electronic health records (EHR) with rich longitudinal survey data, such as the All of Us Research Program, enable computational modeling of stage-specific risk and provide an opportunity to connect predictive factors with potential treatment options.
Results: We developed a multi-stage computational framework to model transitions across OUD onset, treatment-related remission, and relapse, and further link risk factors to treatment and intervention. Using linked EHR and survey data from All of Us, we constructed three time-to-event phenotypes: (i)~OUD onset, defined from opioid prescriptions and diagnostic codes; (ii)~transition to remission; and (iii)~transition to relapse following remission. For each participant, we derived longitudinal predictors from clinical conditions and survey concepts, including recent (1-, 3-, 12-month) event counts, cumulative exposures, and time since last event occurrence. We fit regularized survival models for each transition and aggregated feature selection frequencies and hazard ratios to identify a compact set of high-confidence predictors.
Across all stages, we observed consistent contributions from pain, mental health, and polysubstance use. Chronic pain syndromes, tobacco and nicotine dependence, anxiety and depressive disorders, and cannabis dependence were frequent and prominent predictors of OUD onset and relapse, whereas tobacco dependence during remission and other remission-coded conditions were strongly associated with the transition to remission.
To support computational prioritization of therapeutic strategies, we constructed a therapy knowledge graph integrating genetic targets, biological pathways, and published evidence to link identified risk factors with candidate treatments reported in recent OUD studies and clinical guidelines.
}

\maketitle

\section{Introduction}
The opioid crisis remains a major public health issue, with opioid use disorder (OUD) contributing substantially to preventable morbidity and mortality worldwide \cite{1, 2, 3}. While prescription opioid exposure is a common entry point into OUD, subsequent trajectories are highly heterogeneous, shaped by interacting clinical, behavioral, and social factors. This heterogeneity, coupled with non-monotonic transitions between onset, remission, and relapse over extended follow-up, makes OUD a challenging target for longitudinal risk modeling \cite{8, 9, 10}.

Prior work has demonstrated the utility of electronic health recorders (EHRs) and related real-world data for predicting OUD incidence, overdose, or related adverse outcomes using machine learning and survival-based methods. These approaches have highlighted the predictive value of high-dimensional clinical features derived from diagnoses, medications, and use patterns. However, most studies model OUD as a single terminal event, implicitly collapsing heterogeneous disease trajectories into a binary or time-to-first-event outcome \cite{4, 5, 6, 7}. This limits the ability to distinguish stage-specific risk factors, characterize remission and relapse dynamics, or assess how predictors and interventions operate differently across phases of the disorder \cite{8, 9, 10}. 

Large, diverse, and longitudinal cohorts linking EHRs with patient-reported and behavioral data provide an opportunity to address these limitations and characterize these stages at scale \cite{11, 12}. The All of Us Research Program is particularly well suited for such analyses, as it integrates longitudinal EHR data with survey responses and other data types capturing clinical comorbidities, mental health conditions, substance use, and social and behavioral determinants of health in a diverse population \cite{11, 12, 13}. From a modeling perspective, these data enable the construction of time-varying, multiscale predictors that reflect both recent and cumulative exposures. At the same time, they introduce substantial methodological challenges, including high dimensionality, sparsity, irregular observation times, correlated features, and missing data mechanisms that complicate inference and interpretation \cite{14, 15, 16}. Regularization and structured feature engineering are therefore critical for scalable and interpretable longitudinal modeling.

In parallel, there is a rich but fragmented literature on evidence-based treatments for OUD, including medications for opioid use disorder (MOUD), behavioral therapies, and care delivery models \cite{17, 18, 19, 20}. While these interventions are well characterized clinically, computational methods for systematically linking patient-level risk profiles to therapeutic options remain underdeveloped. Knowledge graphs offer a general framework for integrating heterogeneous biomedical knowledge, including drugs, targets, pathways, and clinical evidence, and for supporting prioritization and interpretability in complex treatment decision spaces \cite{21, 22, 23}. Applying such representations to OUD offers an opportunity to bridge predictive modeling with actionable treatment knowledge.

Here, we present a computational framework that combines regularized survival models of OUD trajectories with an OUD treatment-oriented knowledge graph, and apply it to data from the All of Us Research Program. Our contributions are threefold: (i) a unified modeling framework for three clinically meaningful OUD transitions: onset, remission, and relapse; (ii) a feature-engineering pipeline that identifies compact, recurrent risk factor sets from thousands of candidate EHR- and survey-derived predictors; and (iii) a therapy knowledge graph that links these risk factors to evidence-based OUD treatments curated from recent studies and clinical guidelines, enabling interpretation and computational decision prioritization.

\section{Materials and Methods}
\subsection{Data source and cohort}
We used data from the All of Us Research Program~\cite{24}, a large U.S. cohort that includes electronic health records (EHRs), survey responses, and physical measurements. We limited our analysis to participants with linked EHR data who had at least one record in the Observational Medical Outcomes Partnership (OMOP) tables used in this study (see Supplementary Methods).

\subsection{OUD phenotypes and transitions}

We defined OUD using standard condition and procedure concepts related to opioid use disorder, opioid dependence, and opioid abuse, combined with opioid prescription and medication records (concept sets and code lists are provided in Supplementary Tables). Using these events, we constructed three time-to-event outcomes:

OUD onset: time from cohort entry to the first qualifying OUD event.

Transition to remission: among individuals with OUD, time from the last OUD event to opioid-related remission code, coded as remission.

Transition to relapse: among individuals in remission, time from remission to a subsequent OUD event or return to opioid use.

Individuals were censored at the end of available data.

\subsection{Feature engineering}

We constructed a high-dimensional set of candidate predictors from EHR condition concepts and survey-derived concepts. For each concept, we derived the following time-indexed features relative to the risk-set time $t$:

\begin{itemize}
  \item \texttt{cnt\_30d}, \texttt{cnt\_90d}, \texttt{cnt\_365d}: counts of occurrences in the 30/90/365 days prior to $t$;
  \item \texttt{cum}: cumulative count of occurrences observed up to $t$;
  \item \texttt{days\_since\_last}: the number of days since the most recent recorded occurrence prior to $t$.
\end{itemize}

For example, \texttt{rx\_cnt\_30d} indicates the number of times a specific medication exposure (drug concept) was recorded in the 30 days before $t$; \texttt{mat\_cum} indicates the cumulative number of times a survey concept (e.g., a survey response or attribute concept) was recorded up to $t$; and \texttt{cond\_days\_since\_last} indicates the number of days since the most recent recorded diagnosis/condition concept prior to $t$.

We applied this feature construction scheme to EHR condition concepts (e.g., pain, mental health, substance use, and cardiometabolic disease) and to selected survey questions (e.g., ``About how old were you when you were first told you had X?'') and added basic demographic and clinical covariates including age, sex, family income level, educational attainment, marital status, religious activities, and 20 principle components (PCs) of the genetic relationship matrix.

\subsection{Survival modeling and feature selection}

Because the full feature matrix is extremely high-dimensional and sparse, we used a regularized survival modeling strategy (see Supplementary Methods for details):

We fit separate L1-penalized Cox proportional hazards models (lasso)~\cite{32} for each transition (OUD onset, remission, relapse) using the start--stop (counting-process) formulation. The lasso penalty was applied to all candidate predictors, and the tuning parameter was selected by $K$-fold cross-validation (implemented in \texttt{glmnet}~\cite{33}). For each fitted model, we recorded predictors with non-zero coefficients and their estimated log-hazard coefficients, and summarized effects as hazard ratios ($\mathrm{HR}=\exp(\beta)$). To assess stability across data partitions, we aggregated results across shards by computing each feature's selection frequency and the mean hazard ratio among runs in which the feature was selected.

This procedure yields a table for each transition (onset, remission, relapse) with counts of how often a feature was selected, along with average effect estimates. We used a combination of selection frequency and hazard ratio magnitude to prioritize a compact set of interpretable features for downstream analysis and visualization.

\subsection{Knowledge graph construction}\label{sec:kg}

To connect risk factors to treatments, we developed a therapy knowledge graph from recent OUD genetic studies and relevant clinical resources~\cite{25,26}, focusing on medications for opioid use disorder (MOUD; e.g., buprenorphine, methadone, and naltrexone).

We constructed an OUD-relevant gene set by harmonizing evidence from large-scale GWAS and meta-analyses of OUD and related substance use disorders~\cite{25,26}.
Gene--pathway relationships were obtained from Reactome~\cite{27,28}, and drug--target links and drug annotations were compiled from ChEMBL, RxNorm, and DrugBank~\cite{29,30,31}.

\paragraph{Graph schema.}
We build a directed, weighted graph $G=(V,E)$ with nodes
$V=\{\mathrm{OUD}\}\cup \mathcal{G}\cup \mathcal{P}\cup \mathcal{D}$ representing the disease (OUD), genes ($\mathcal{G}$), pathways ($\mathcal{P}$), and drugs ($\mathcal{D}$).
Edge types include:
(i) $\mathrm{OUD}\rightarrow g$ (\textsc{HAS\_GENE} and \textsc{HAS\_GENE\_BRIDGED}),
(ii) $g\rightarrow d$ (\textsc{TARGETED\_BY}), and
(iii) $g\rightarrow p$ (\textsc{IN\_PATHWAY}), used to support bridged $\mathrm{OUD}\rightarrow g$ edges.

\paragraph{Evidence-weighted OUD$\rightarrow$Gene edges (direct).}
For each seed gene $g\in \mathcal{G}_{\mathrm{seed}}$, we compute an association weight $a_g$ from GWAS evidence:
\begin{equation}
a_g=
\begin{cases}
0.5 + 0.3\, r_{\mathrm{rep},g}, & \text{if replication count is available},\\
1.0, & \text{if flagged in MVP cross-ancestry study\cite{25}},\\
0.8, & \text{if flagged in OUD-only study\cite{26}},\\
0.5, & \text{if flagged in OUD-only MTAG\cite{26}},\\
0.2 + \mathrm{posterior}_g, & \text{if posterior is available},\\
\log(3 + p_{\mathrm{rank},g})/e, & \text{if $p$-value rank is available},\\
\varepsilon, & \text{otherwise}.
\end{cases}
\label{eq:ag}
\end{equation}
We cap the direct disease--gene weight:
\begin{equation}
w^{\mathrm{direct}}_{\mathrm{OUD}\rightarrow g}=
\max\!\left(\varepsilon,\ \min(a_g,\ c_{\max})\right).
\label{eq:w_direct}
\end{equation}

\paragraph{Pathway-bridged OUD$\rightarrow$Gene edges.}
Let $M_p$ denote the member genes of pathway $p$.
If $p$ contains at least $m_{\min}$ direct seed genes $S_p = M_p \cap \mathcal{G}_{\mathrm{seed}}$, we bridge other pathway members using a $\gamma$-penalized bonus:
\begin{equation}
b_p=\gamma \cdot \frac{\sum_{g\in S_p} w^{\mathrm{direct}}_{\mathrm{OUD}\rightarrow g}}{|M_p|}.
\label{eq:bp}
\end{equation}
For each non-seed member $h \in M_p \setminus \mathcal{G}_{\mathrm{seed}}$,
\begin{equation}
w^{\mathrm{bridged}}_{\mathrm{OUD}\rightarrow h} \leftarrow b_p.
\label{eq:w_bridged}
\end{equation}
We cap the total number of bridged genes by $B_{\max}$ and record supporting pathways for explainability.

\paragraph{Gene$\rightarrow$Drug target edges.}
To avoid overfull equations in two-column format, we use compact indicators:
$I^{\mathrm{ant}}_{g,d}$ (antagonist/inhibitor), $I^{\mathrm{ago}}_{g,d}$ (agonist/activator),
$I^{\mathrm{ph3}}_{d}$ (phase $\ge 3$), $I^{\mathrm{app}}_{d}$ (approved),
$I^{\mathrm{bbb}}_{d}$ (BBB-permeable), and $I^{\mathrm{res}}_{d}$ (research-only).
For a gene--drug pair $(g,d)$, we compute:
\begin{equation}
\operatorname{clip}(x,0,1)=\min\{1,\max\{0,x\}\}.
\end{equation}
\begin{equation}
\begin{aligned}
t_{g\rightarrow d} \;=\;& 1
+ \operatorname{clip}\!\left(\frac{p\mathrm{ChEMBL}_{g,d}-5}{4},\,0,\,1\right)
+ 0.2\, I^{\mathrm{ant}}_{g,d}
+ 0.1\, I^{\mathrm{ago}}_{g,d} \\
&+ 0.3\, I^{\mathrm{ph3}}_{d}
+ 0.7\, I^{\mathrm{app}}_{d}
+ 0.2\, I^{\mathrm{bbb}}_{d}
- 0.5\, I^{\mathrm{res}}_{d}.
\end{aligned}
\label{eq:tgd}
\end{equation}
and set:
\begin{equation}
w_{g\rightarrow d}=\max(\varepsilon,\, t_{g\rightarrow d}).
\label{eq:w_gd}
\end{equation}

\paragraph{Row normalization (stochastic transitions).}
All outgoing weights are row-normalized to form a transition matrix $P$:
\begin{equation}
P_{ij}=\frac{w_{ij}}{\sum_k w_{ik}}.
\label{eq:row_norm}
\end{equation}

\paragraph{Ranking via Personalized PageRank (PPR)\cite{34, 35}.}
With personalization vector $s$ centered on OUD (optionally including an approved-drug prior), we compute:
\begin{equation}
r = \alpha P^{\mathsf{T}} r + (1 - \alpha)s,
\label{eq:ppr}
\end{equation}
and rank drugs by $S(d)=r(d)$ (or by $S(d)$ plus optional node-level priors when used).

\paragraph{Outputs.}
We write a ranked drug table (TSV; \texttt{--out}) with:
\begin{itemize}
  \item \texttt{rank}, \texttt{drug\_name/id}, score $S(d)$;
  \item supporting OUD genes and gene labels (direct/bridged/non-seed);
  \item provenance fields (e.g., supporting genes/pathways) and key annotations (approved, BBB).
\end{itemize}

\paragraph{Pseudocode.}
\vspace{-0.3em}
\noindent\begin{minipage}{\columnwidth}
\captionof{algorithm}{Knowledge graph construction and drug ranking}\label{alg:kg}
\footnotesize
\setlength{\tabcolsep}{0pt}
\begin{algorithmic}[1]
\State \textbf{Input:} nodes, edges, metadata
\State $g \gets \textsc{build\_graph}(\text{nodes}, \text{edges}, \gamma, B_{\max}, m_{\min}, \text{alias\_mode}, \text{meta})$
\State $sg \gets \textsc{project\_disease\_gene\_drug}(g)$
\State \textsc{normalize\_outgoing\_weights}$(sg)$ \Comment{make $P$ row-stochastic}
\State $s \gets \textsc{build\_personalization}(sg,\ \text{src}=\text{``OUD''},\ \text{approved\_prior})$
\State $r \gets \textsc{pagerank}(sg,\ \alpha,\ s,\ \text{weight}=\text{``weight''})$
\Statex \hspace{1.2em}$\text{rows} \gets \textsc{rank\_with\_why}(sg,\ r,\ \text{alias\_mode},\ \text{borrow},\ \text{why\_top},\ \text{why\_out})$
\State \textsc{write\_tsv}(\texttt{--out}, rows)
\end{algorithmic}
\end{minipage}
\vspace{-0.6em}

\section{Results}
\subsection{Cohort characteristics}\label{sec:cohort_characteristics}
The analytic cohort included $N=214{,}387$ participants. The mean age was 48.26 years (SD 16.21). Ages were right-skewed, with an interquartile range (IQR) of 35.33--60.65 years and a maximum of 103.03 years.

Regarding sex at birth, 62.46\% of participants were female and 36.48\% were male; the remaining 1.06\% reported intersex, ``prefer not to answer,'' or had missing values. The racial composition was predominantly White (56.59\%), followed by Black or African American (17.19\%) and Asian (2.01\%), with additional categories (e.g., Multiple and American Indian or Alaska Native) comprising smaller proportions.

\subsection{Predictors of OUD onset}\label{sec:predictors_onset}
\begin{figure}[t]
\centering
\includegraphics[width=\columnwidth]{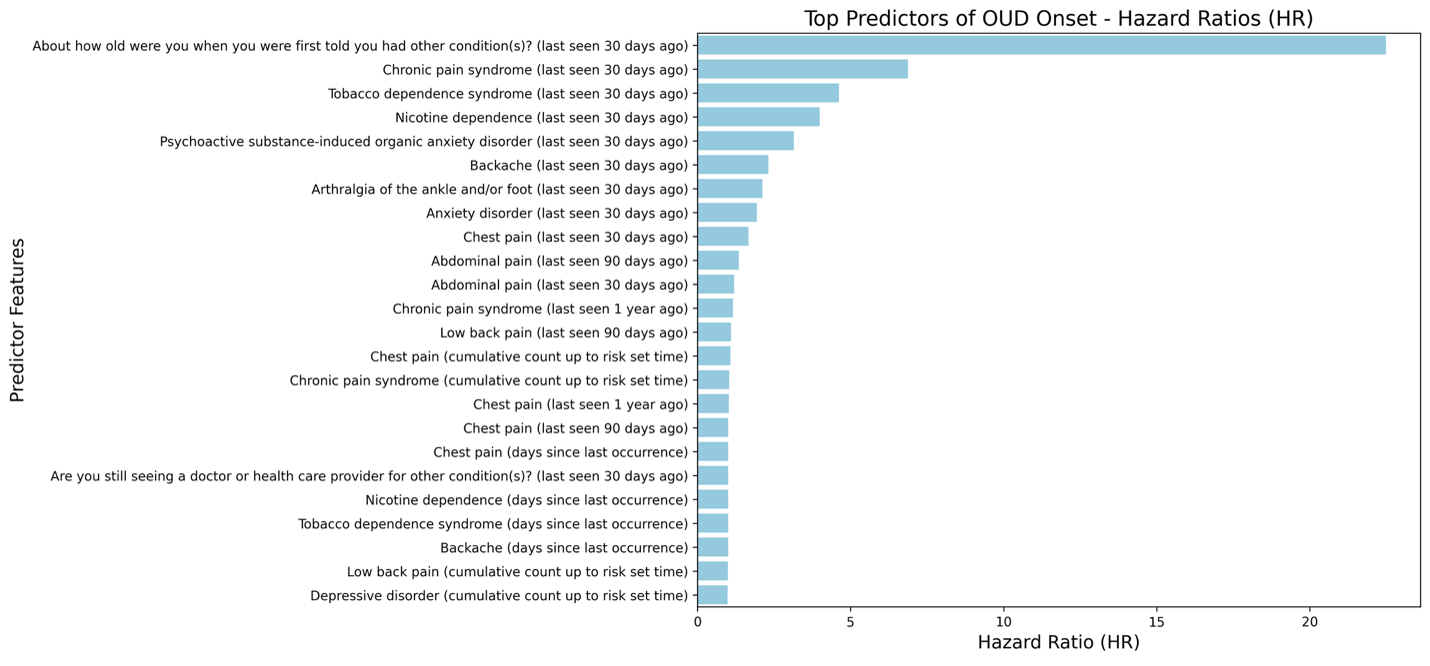}
\caption{\textbf{Predictors of OUD onset.} Hazard ratios (HRs) from time-to-event models for OUD onset. HR$>1$ indicates increased hazard (earlier onset) and HR$<1$ indicates decreased hazard. Predictors shown are the 24 prioritized features selected by the modeling/feature-selection pipeline; \textit{Selected} indicates recurrence across resampling/validation runs. Abbreviations: OUD, opioid use disorder; EHR, electronic health record; HR, hazard ratio.}
\label{fig:onset_hr}
\end{figure}

Among the 24 prioritized predictors for OUD onset, we observed strong contributions from pain-related conditions, tobacco/nicotine use, and mental health (Figure~\ref{fig:onset_hr}).

EHR-derived features capturing chronic pain syndrome, back pain, chest pain, and low back pain were associated with elevated hazards of OUD onset. Tobacco dependence syndrome and nicotine dependence were consistently selected and were associated with increased risk. Anxiety disorder and depressive disorder also appeared as recurrent predictors. In addition, survey items such as ``About how old were you when you were first told you had other condition(s)?'' were selected with high hazard ratios, suggesting that earlier and more complex medical histories may mark individuals at greater risk. Together, these results highlight a pattern of pain--mental health--tobacco comorbidity as a key substrate for OUD onset risk.

\subsection{Predictors of OUD remission}\label{sec:predictors_remission}

\begin{figure}[t]
\centering
\includegraphics[width=\columnwidth]{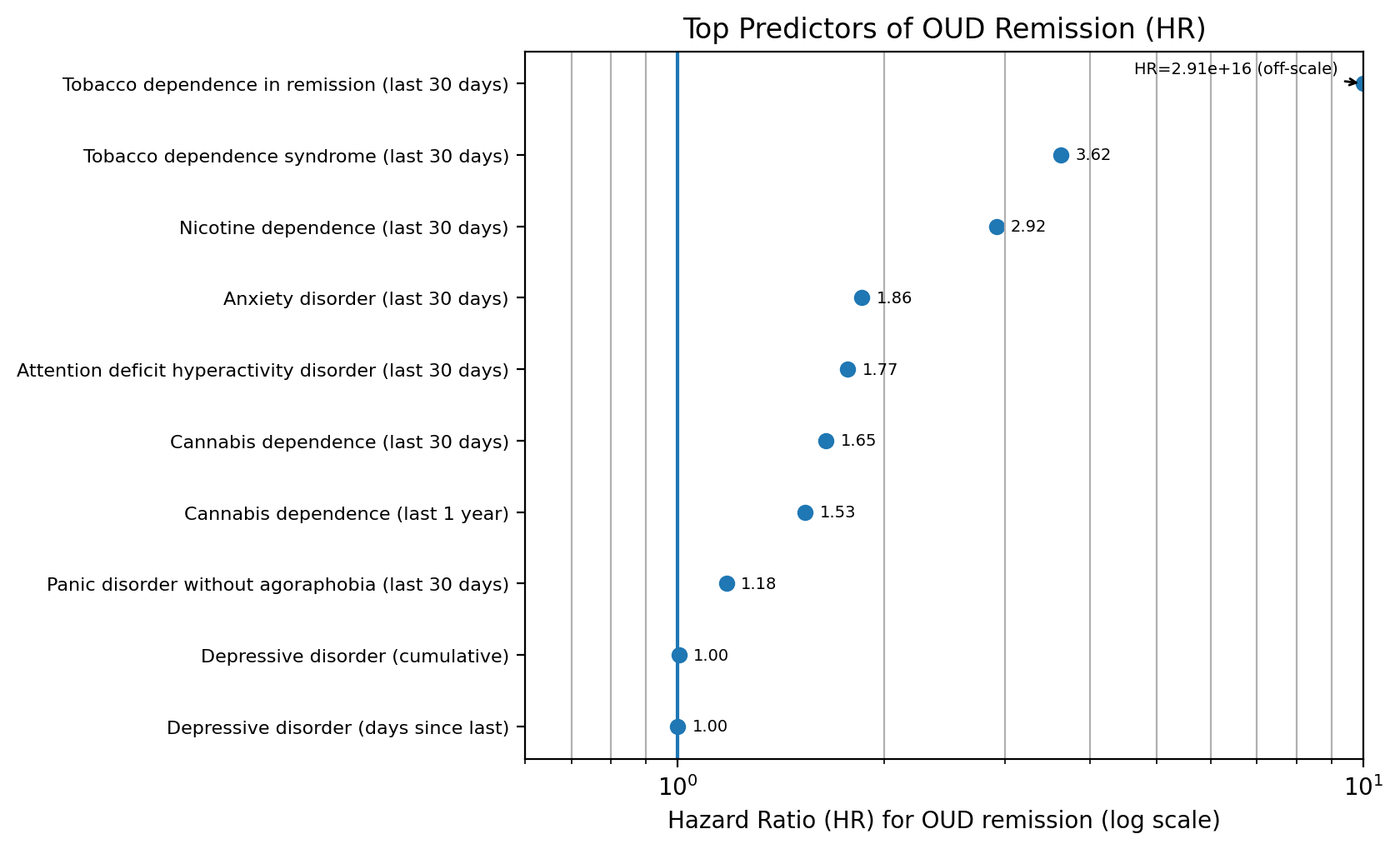}
\caption{\textbf{Top predictors of OUD remission.} Hazard ratios (HRs) summarize associations with time to OUD remission; HR$>1$ indicates a higher hazard of remission (i.e., faster remission), and HR$<1$ indicates a lower hazard. To improve readability, the x-axis is shown on a log scale; values above 10 are plotted at the axis limit and annotated as off-scale. Predictors shown are the 10 prioritized features in the remission model, including recent-condition counts (last 30 days; last 1 year) and longitudinal summaries (cumulative counts; days since last occurrence). Abbreviations: OUD, opioid use disorder; HR, hazard ratio.}
\label{fig:remission_hr}
\end{figure}

In the remission models, we prioritized 10 predictors, dominated by substance-use and anxiety-related conditions (Figure~\ref{fig:remission_hr}).

Tobacco dependence in remission (last 30 days) yielded an extremely large HR estimate ($\approx 2.91\times 10^{16}$), which may reflect sparse data and/or separation, or close temporal alignment with remission-related care; this estimate should be interpreted cautiously. Among the remaining predictors, tobacco dependence syndrome (HR$=3.62$) and nicotine dependence (HR$=2.92$) were consistently associated with higher remission hazards, and anxiety disorder (HR$=1.86$) showed a moderate positive association. Cannabis dependence appeared in both a recent window (last 30 days; HR$=1.65$) and a longer window (last 1 year; HR$=1.53$), while panic disorder without agoraphobia had a smaller effect (HR$=1.18$). Depressive-disorder summaries were near null (HR$\approx 1.00$).

Together, these patterns emphasize the importance of substance-use comorbidity and anxiety-related conditions during the transition from active OUD to remission, potentially capturing variation in clinical engagement, care trajectories, and psychiatric burden during recovery.

\subsection{Predictors of relapse}\label{sec:predictors_relapse}

\begin{figure}[t]
\centering
\includegraphics[width=\columnwidth]{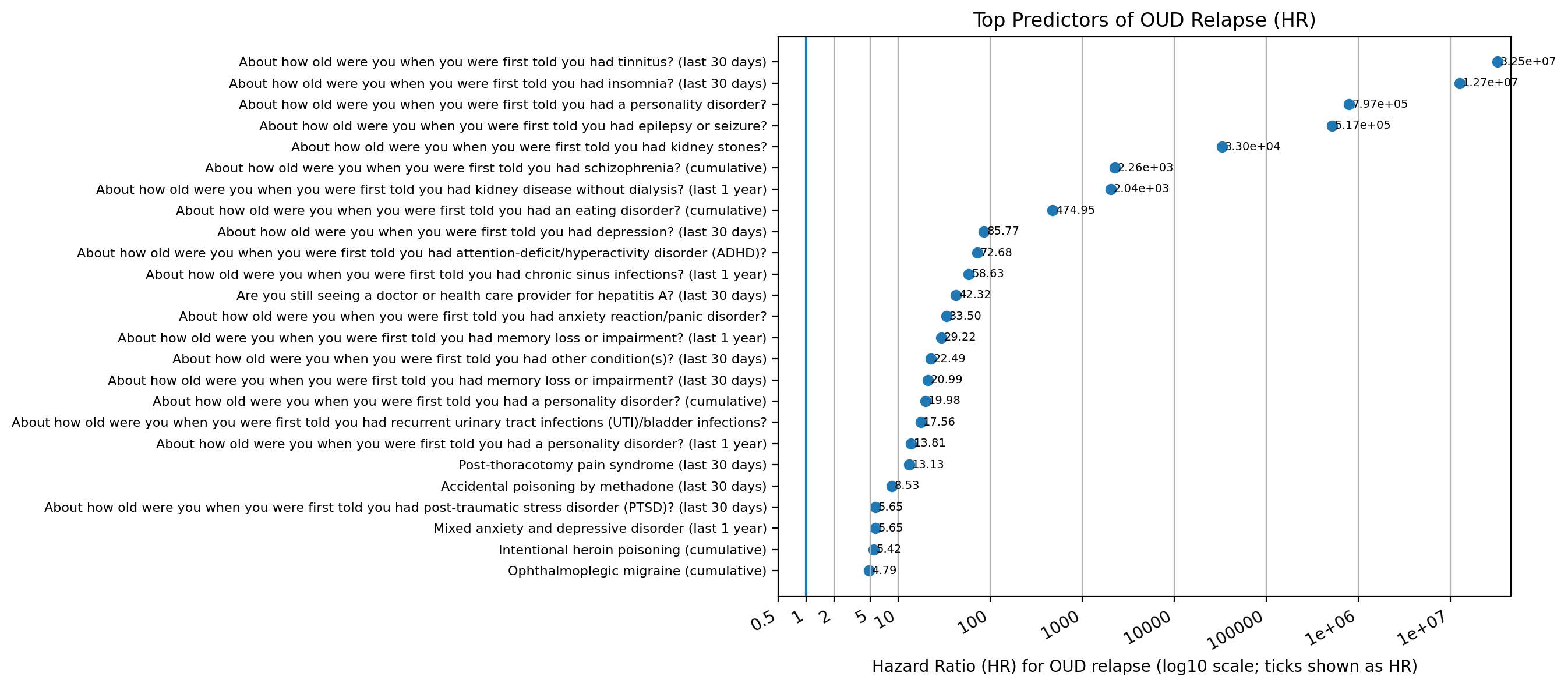}
\caption{\textbf{Top predictors of OUD relapse (hazard-ratio point estimates).} Hazard ratios (HRs) summarize point-estimate associations with time to relapse after remission; HR$>1$ indicates a higher hazard of relapse. Predictors shown are the top-ranked features from the relapse screening/prioritization pipeline. The x-axis is log-scaled to accommodate extreme HR values driven by sparse counts.}
\label{fig:relapse_hr}
\end{figure}

For relapse, we screened 247 candidate predictors and identified a prioritized set, with substantial overlap with the onset model but several notable differences (Figure~\ref{fig:relapse_hr}). The top-ranked predictors were dominated by survey-based ``age-at-diagnosis'' items spanning diverse conditions (e.g., sleep-related disorders, renal conditions, chronic infections, fatigue, and personality-related diagnoses). Many of these features yielded extreme hazard-ratio point estimates, consistent with sparse counts and/or strong correlations with broader medical complexity. In addition, conditions related to chronic pain, mood/anxiety disorders, depressive disorders, and polysubstance involvement (e.g., cannabis dependence; cocaine dependence in remission) were repeatedly prioritized. Because confidence intervals and $p$-values are not shown for this screening-based ranking, these associations should be interpreted as hypothesis-generating indicators of clinical complexity and comorbidity patterns among individuals at higher relapse risk after remission.

Together, these patterns emphasize the importance of substance-use comorbidity and anxiety-related conditions during the transition from active OUD to remission, potentially capturing variation in clinical engagement, care trajectories, and psychiatric burden during recovery.

\subsection{Significant summary features from the OUD survival models}\label{sec:summary_features}

We examined a focused set of summary ``activity'' features capturing recent and cumulative healthcare utilization and treatment exposure. These included counts of prescriptions (\texttt{rx}) and diagnoses/conditions (\texttt{cond}) over recent windows (last 30/90/365 days) and cumulatively, as well as medication-assisted treatment (MAT/MOUD) summaries. Given the large person-period dataset, several of these features remained statistically significant after multiple-testing correction (FDR $<0.05$). Full results, including effect sizes and uncertainty (HR, 95\% CI, and FDR-adjusted $p$-values), are reported in the Supplementary Information.

\subsubsection{OUD onset (developing OUD)}
In the onset model, recent clinical activity was consistently associated with a higher hazard of developing OUD (Supplementary Table S6, Figure S4). Specifically, higher counts of prescriptions and diagnoses in the last 30 and 90 days had hazard ratios slightly above 1.0. Although these per-unit effects are modest (predictors are modeled per unit of $\log(1+\mathrm{count})$), they are estimated with high precision due to the large sample size. Overall, this pattern suggests that periods of increased healthcare activity---reflected by more prescriptions and recorded conditions---co-occur with greater short-term risk of OUD onset.

\subsubsection{Recovery (remission from OUD)}
In the recovery model , MAT/MOUD intensity emerged as a dominant predictor set (Supplementary Table S7, Figure S5). Greater MAT exposure over the last 30/90/365 days and greater cumulative MAT history were associated with a higher hazard of recovery (HRs $>1$), whereas longer time since the last MAT event was associated with a lower hazard of recovery (HR $<1$). This pattern is clinically plausible: individuals actively receiving MAT and those with more recent MAT contact tend to remit sooner. We also observed positive associations for recent condition and prescription counts, which may reflect clinical engagement and closer follow-up, where improvement and remission are more likely to be observed and documented. Taken together, the recovery results suggest that ongoing treatment exposure and clinical engagement track with earlier remission.

\subsubsection{Relapse (return to OUD after remission)}
\begin{figure}[t]
\centering
\includegraphics[width=\columnwidth]{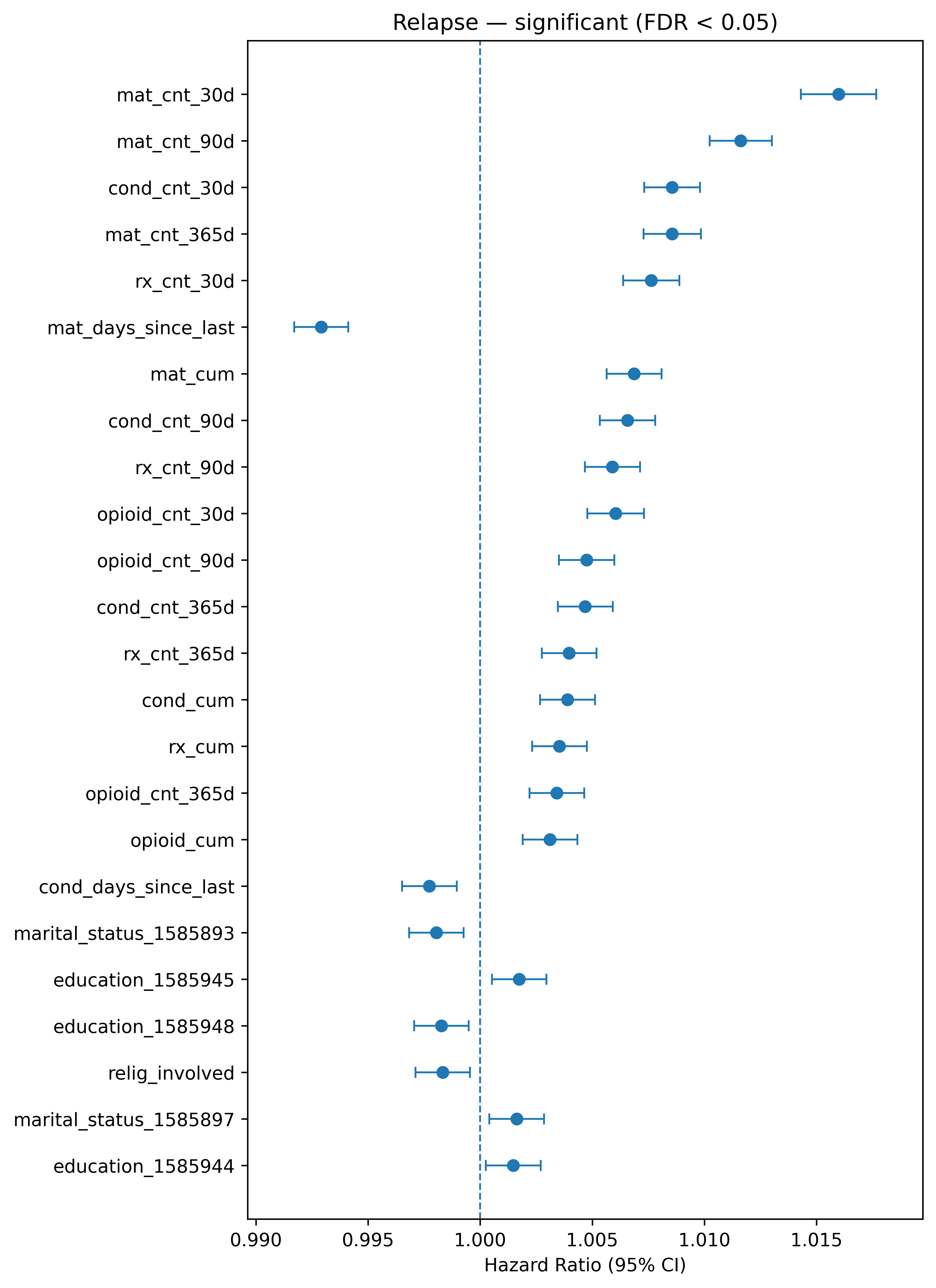}
\caption{\textbf{Relapse model significant predictors (FDR $<0.05$).} Forest plot of adjusted hazard ratios (HR) and 95\% confidence intervals (CIs) from the relapse transition model. Points denote HR estimates and horizontal bars indicate 95\% CIs; the dashed vertical line marks $\mathrm{HR}=1$. Predictors are modeled per unit increase in $\log(1+x)$-transformed counts for utilization/count features (e.g., MAT, prescriptions, diagnoses, opioid-related counts) and per unit increase in “days since last” variables. Only predictors passing FDR $<0.05$ are shown. The case indicator is excluded because it encodes cohort labeling rather than a clinical predictor.}
\label{fig:relapse}
\end{figure}

In the relapse model (Supplementary Table S8, Figure S6), relapse hazard was associated with markers of recent and cumulative clinical activity, including higher MAT counts, prescription counts, diagnosis counts, and opioid-related counts over 30/90/365-day windows and cumulatively (HRs modestly $>1$ per $\log(1+\mathrm{count})$ unit) ) (Figure~\ref{fig:relapse}). These associations are likely influenced by confounding by severity and surveillance/ascertainment: individuals with greater clinical instability or higher underlying risk are more likely to have frequent visits, more prescriptions, continued MAT adjustments, and more opportunities for relapse to be detected and coded. Consistent with this interpretation, time since last MAT and time since last condition had HRs slightly below 1, indicating that relapse risk is highest during clinically active periods and decreases as time elapses since the most recent treatment or diagnosis event. A small number of baseline demographic indicators (e.g., education or marital status categories) reached statistical significance, but effect sizes were extremely small, suggesting weak associations and/or proxies for unmeasured factors. Overall, the relapse model indicates that relapse is concentrated in periods of high clinical activity and higher illness burden, and that MAT exposure in this context likely functions primarily as a marker of severity/engagement rather than a direct cause of relapse.

\subsubsection{Summary interpretation}
Across the three transitions (onset, recovery, and relapse), a consistent theme is that recent healthcare activity and treatment exposure are informative predictors of event timing. For onset, increased recent activity aligns with elevated hazard of developing OUD. For recovery, active and recent MAT exposure aligns with earlier remission. For relapse, increased activity and opioid/MAT exposure align with higher observed relapse hazard, plausibly reflecting severity and increased observation during clinically unstable periods. Because effect sizes are small but precisely estimated, we emphasize interpretation in terms of risk tracking and care engagement rather than causal claims.

\section{Knowledge graph integration}\label{sec:kg_integration}

\begin{figure}[t]
\centering
\includegraphics[width=\columnwidth]{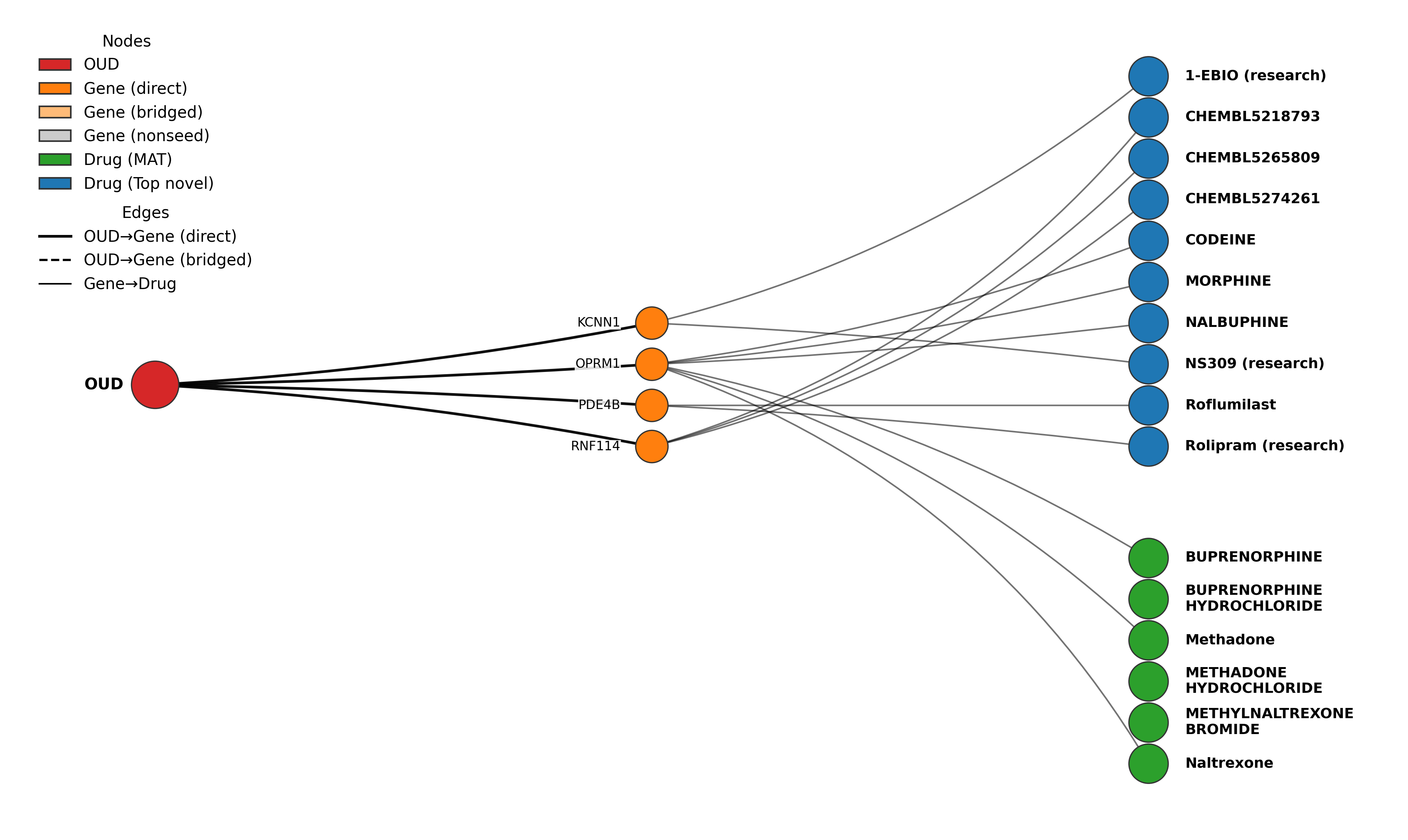}
\caption{\textbf{Knowledge graph drug repurposing results for OUD.} The graph connects OUD (red) to GWAS/meta-analysis prioritized genes (orange; including direct seeds and pathway-bridged genes) and to drugs via curated gene--drug target relationships. Green nodes indicate established medication-assisted treatments (MAT/MOUD) for OUD; blue nodes indicate non-MAT drugs prioritized by Personalized PageRank starting from the OUD node. Recovery of known MAT/MOUD drugs provides face validity. Non-MAT candidates cluster around mechanistic axes such as opioid receptor signaling (\textit{OPRM1}-linked), PDE4/cAMP signaling (\textit{PDE4B}-linked), and ion-channel modulation (\textit{KCNN1}-linked), suggesting hypotheses for adjunctive or alternative therapeutic exploration. Duplicate drug entries reflect ingredient versus salt forms in source databases.}
\label{fig:kg_results}
\end{figure}

Figure~\ref{fig:kg_results} highlights two key findings from the knowledge-graph analysis. First, the graph passes an important sanity check by prioritizing medications already used to treat OUD. Buprenorphine, methadone, and naltrexone appear prominently and connect to OUD through well-established biology such as the opioid receptor gene \textit{OPRM1}. This recovery of established MAT/MOUD drugs supports the face validity of the graph and suggests that the same evidence-weighted scoring can be used to propose additional candidates.

Second, the top-ranked non-MAT drugs cluster into a small number of interpretable mechanistic groups based on the OUD-linked genes they target. Some candidates connect through \textit{OPRM1} and related opioid-receptor pharmacology; we treat these primarily as confirmation that the graph highlights core OUD biology, while noting that certain opioid agonists are not appropriate repurposing options because of abuse liability and should be filtered out. Other candidates connect through \textit{PDE4B} (cAMP/PDE4 signaling; e.g., roflumilast or rolipram) or through \textit{KCNN1} (ion-channel modulation; e.g., NS309 or 1-EBIO), suggesting non-opioid mechanisms that may be explored as adjunctive or alternative strategies. Because the ranking reflects network proximity and evidence-weighted connectivity rather than causal efficacy, we interpret these results as hypothesis-generating. In downstream use, candidates can be refined using practical filters (e.g., approval status, safety/abuse potential, CNS relevance, and feasibility) and then evaluated using independent evidence (literature review, EHR-based analyses, or preclinical studies). Duplicate drug entries largely reflect database normalization (ingredient versus salt forms) rather than distinct mechanisms and can be merged in the final ranked outputs (Supplementary Information).

\section{Discussion}\label{sec:discussion}

We present a multi-stage framework for modeling OUD trajectories---onset, remission, and relapse---in the All of Us Research Program, demonstrating how high-dimensional EHR and survey features can be distilled into compact sets of interpretable predictors. Across stages, we repeatedly observed the joint roles of chronic pain, mental health disorders, and polysubstance use, particularly tobacco and cannabis, in shaping OUD trajectories.

Our survival modeling strategy is scalable to large cohorts and robust to sparsity, and the resulting feature-frequency summaries are directly interpretable by clinicians and domain experts. By integrating these summaries with an OUD treatment knowledge graph, we move beyond risk prediction alone toward a framework that can support clinical decision support and hypothesis generation about targeted interventions.

The multi-state survival models also highlighted consistent associations between recent clinical activity and transition timing. Higher recent clinical activity (e.g., diagnosis counts and opioid-containing prescription summaries) was associated with a modestly increased hazard of OUD onset (e.g., HR per one $\ln$-unit increase in 30-day diagnosis count $\approx 1.003$, FDR$<0.05$). Greater exposure to medications for OUD predicted faster recovery (HR $\approx 1.01$--$1.02$ per $\ln$-unit; all FDR$<0.05$), whereas longer time since the last OUD treatment event predicted slower recovery (HR $\approx 0.992$, FDR$<0.05$), consistent with remission co-occurring with recent treatment engagement. Finally, higher counts of recent OUD treatment, opioid prescriptions, and diagnoses were associated with increased relapse hazard (HR $\approx 1.006$--$1.016$ per $\ln$-unit; FDR$<0.05$). These associations are likely influenced by confounding due to severity and surveillance/ascertainment and therefore should not be interpreted as evidence of causal effects.

Several limitations should be noted. First, our current screening and prioritization procedures rely on feature-selection frequency and average hazard ratios rather than fully propagated uncertainty for each feature; future work will compute stage-specific confidence intervals and $p$-values for a refined subset of predictors. Second, remission and relapse definitions derived from EHR data are imperfect and may misclassify individuals with intermittent treatment, care received outside captured systems, or incomplete documentation. Third, the therapy knowledge graph currently covers a limited set of studies and guidelines; expanding it to incorporate additional trials, implementation studies, and updated clinical recommendations is an important next step.

Despite these limitations, our results show that large-scale cohorts such as All of Us can support fine-grained, stage-specific risk modeling of OUD, and that combining these models with structured treatment knowledge offers a promising path toward personalized and interpretable decision support in addiction medicine.

\section{Conclusion}

Our work illustrates how survival modeling can be combined to characterize the complex trajectories of opioid use disorder. By identifying stage-specific patterns of comorbid pain, mental health conditions, and polysubstance use, and linking them to evidence-based treatment strategies, we provide a foundation for developing transparent, clinically meaningful decision support tools for OUD prevention and care.

\section{Conflicts of interest}
The authors declare no competing interests.

\section{Funding}

This work was supported by the National Institutes of Health (NIH), National Institute on Drug Abuse (NIDA) under award DP1DA054373. The funder had no role in the study design; data collection, analysis, or interpretation; manuscript writing; or the decision to submit for publication. The content is solely the responsibility of the authors and does not necessarily represent the official views of the NIH.

\section{Author Contributions}
Mengman Wei conceived the study, designed the analytical framework, performed all data processing, statistical analyses, and computational modeling, and drafted the manuscript. All code implementation, data curation, and result interpretation were conducted by Mengman Wei.\\

Qian Peng provided supervision, general guidance, resource support, and funding acquisition.

\section{Preprint Notice}

This manuscript is a preprint and has not yet undergone peer review. The content is shared to disseminate findings and establish precedence. Additional analyses and revisions may be incorporated in future versions.

\section{Data availability}
\textbf{Code.} The analysis code and scripts used in this study are available at: \url{https://github.com/mw742/OUD-survival}

\textbf{Data.} This study uses data from the NIH \textit{All of Us} Research Program (Researcher Workbench; controlled tier). Access is subject to program policies and approval.

\bibliographystyle{unsrt}
\bibliography{reference}




\end{document}